\documentclass[11pt]{article} 

\usepackage[utf8]{inputenc} 

\usepackage{geometry} 
\usepackage{graphicx} 

\usepackage{hyperref}
\usepackage{graphicx} 

\usepackage{bm}
\usepackage{graphicx}
\usepackage{caption}
\usepackage{color} 
\usepackage{lineno}
\usepackage{epstopdf}
\usepackage{amsmath} 
\usepackage{animate}
\usepackage{breakcites}
\usepackage{amsfonts}
\usepackage[affil-it]{authblk}
\usepackage{booktabs} 
\usepackage{array} 
\usepackage{paralist} 
\usepackage{verbatim} 
\usepackage{subfig} 

\usepackage{fancyhdr} 
\usepackage{sectsty}
\allsectionsfont{\sffamily\mdseries\upshape} 

\usepackage[nottoc,notlof,notlot]{tocbibind} 
\usepackage[titles,subfigure]{tocloft} 

\title{}

\author[1,2*]{Wilten Nicola} \author[3]{Sue Ann Campbell}

\affil[1]{University of Calgary, Cumming School of Medicine, Department of Cell Biology and Anatomy}
\affil[*]{%
  Corresponding Author:  wilten.nicola@ucalgary.ca}
\affil[2]{Hotchkiss Brain Institute}
\affil[3]{University of Waterloo, Department of Applied Mathematics
and Centre for Theoretical Neuroscience}

\title{Normalized Connectomes Show Increased Synchronizability with Age Through Their Second Largest Eigenvalue} 
\date{\today} 

\begin{document}

\maketitle

\begin{abstract}
The synchronization of different brain regions is widely observed under both normal and pathological conditions such as epilepsy.  However, the relationship between the dynamics of these brain regions, the connectivity between them, and the ability to synchronize remains an open question.  We investigated the problem of inter-region synchronization in networks of Wilson-Cowan/Neural field equations with homeostatic plasticity, each of which acts as a model for an isolated brain region.  We considered arbitrary connection profiles with only one constraint: the rows of the connection matrices are all identically normalized. We found that these systems often synchronize to the solution obtained from a single, self-coupled neural region.  We analyze the stability of this solution through a straightforward modification of the Master Stability Function (MSF) approach and found that synchronized solutions lose stability for connectivity matrices when the second largest positive eigenvalue is sufficiently large, for values of the global coupling parameter that are not too large.  This result was numerically confirmed for ring systems and lattices and was also robust to small amounts of heterogeneity in the homeostatic set points in each node.  Finally, we tested this result on connectomes obtained from 196 subjects over a broad age range (4-85 years) from the Human Connectome Project.  We found that the second largest eigenvalue tended to decrease with age, indicating an increase in synchronizability that may be related to the increased prevalence of epilepsy with old age.   
\end{abstract}

\tableofcontents
\clearpage
\section{Introduction} 

Neuroscience has gained the ability to non-invasively map the structural and functional connectivity of the brain in a diverse group of subjects with the emergence of large scale connectomics initiatives \cite{HCP1,HCP2}.   This capability has created a cornucopia of neuroimaging data in the form of connection matrices or connectomes. These connectomes determine how brain regions are coupled either structurally via white matter tracts, or functionally, through correlations.   However, a natural problem emerges with this glut of data: How do we use it? 
 
The initial approach in analyzing these connectomes originates from graph theory
with the nodes of the network corresponding to the vertices of a mathematical
graph and the elements of the connection matrix determining the edges of the graph \cite{graphtheory}. 
The connectomes are then analyzed for graph-based statistics such as the connectivity, mean in/out degrees, average shortest path length, average motif strengths and frequencies, or community structures, to name a few \cite{graphtheory2}. Typically, these statistics are average properties of the connections between individual nodes or small populations of nodes over the entire graph, although some may be more global in nature.   These graph-based statistics have proven useful for summarizing and elucidating the differences between connectomes of healthy individuals, and those with pathological conditions \cite{graphtheory2}. 

However, the dynamics of a large network are not solely determined by graph based statistics.  Indeed, dynamical systems theory tells us that the eigenvalues of the connection matrix are often the critical determining factor of the stability of any particular solution \cite{eigvalue1}.   This statement was clarified further in \cite{MSF} through the derivation of the so-called Master Stability Function (MSF).  The MSF determines under what conditions a dynamical system synchronizes and only depends on the eigenvalues of a connection matrix, rather than any explicit graph-based measure. 

Given the power of the MSF approach, computational neuroscience has recently started analyzing connectome eigenvalues rather than graph-based statistics \cite{DB1}.  This is typically done by taking the raw structural connectome and computing the Laplacian matrix.  This is a critical step required to apply MSF analysis.  Owing to its construction, the Laplacian effectively assumes that the coupling between nodes is diffusive with strong recurrent coupling within a node and weak coupling between nodes \cite{DB1,MSF2}.  Unfortunately, the evidence for diffusive coupling is scant \cite{BSF1}.    

Here, we consider an alternate assumption on the Diffusion Tensor Imaging (DTI) derived structural connectome that may have some empirical support.  Rather than assuming diffusive connectivity, we instead assume that each neural region has
the same level of incoming activity.  Biologically, this could be achieved by the growth or pruning of dendritic spines at synapses, which would adjust the synaptic weights \cite{segal2000dendritic,bonhoeffer2002spine}.  
In our theoretical work, we achieve this through a normalization process,
where the incoming weights for each node are divided by the total weight 
sum for that node.  
Like the diffusive coupling assumption, this normalization 
leads to an analytically tractable criterion for synchronization through a MSF approach.  We test our approach on a network of Wilson-Cowan nodes with intranode inhibitory homeostatic plasticity \cite{hell1,wiltenchaos,vogels} and various connectivity matrices.  We find that the magnitude of the second largest eigenvalue of the normalized connectivity matrix is the critical determining factor for the stability of the synchronized state, while the largest eigenvalue is identical for all connectomes due to normalization.  
We then apply this approach to DTI derived structural connectome data from the human connectome project for $n=196$ subjects covering a large age group (4-85 years).  We find that the second largest eigenvalue (SLE) of these normalized connectomes decreases with age. Extrapolating from our analysis, this implies increasing synchronizability with age.  This partially mirrors the epidemiology of epilepsy where prevalence increases with old age \cite{epilepsy1,epilepsy2}, indicating a potential diagnostic use of the eigenvalues of the DTI derived structural connectomes.  

\section{Results} 
In order to apply the MSF approach, an explicit model of the dynamics of the nodes and coupling between nodes is necessary.   Here, we consider a modification of the Wilson-Cowan system \cite{wc1} first developed in \cite{hell1,wiltenchaos}.  This system is described by three dynamical variables:
\begin{eqnarray}
\tau_1 \frac{dE_k}{dt} &=& -E_k + \phi\left(\sum_{j=1}^NW^{EE}_{kj} E_j - W^{IE}_k I_k\right)\label{ns1}\\
 \frac{dI_k}{dt} &=& -I_k + \phi\left( W^{IE} I_k \right)\label{ns2}\\
\tau_2 \frac{dW^{IE}_k}{dt} &=& I_k(E_k-p), \quad k=1,2,\ldots N \label{ns3}
\end{eqnarray}
where $E_k$ is the activity of the excitatory population of neurons within the $k$th node, $I_k$ is the activity of the inhibitory population in the $k$th node (Figure \ref{F0}A), and $W^{EI}_k$ is the homeostatically adjusted inhibitory weight.  The population activities are variables that are confined to the interval $(0,1)$.  This loosely corresponds to the proportion of neurons active in node $k$, and are thus ``neural-field" or ``mean-field" approximations to a large population of neurons in node $k$ commonly. These are commonly used in computational neuroscience to investigate diverse network level phenomenon \cite{mf1,mf2,mf3,nfcoombes,mfbreakspear,pb2,be,park}.   The inhibitory homeostatic weight $W^{EI}$ drives the excitatory population activity $E_k$ towards the homeostatic set-point $p$.  Further, we assume that inhibition is always a local interaction within a node while excitation is global and dependent on the coupling matrix $\bm W^{EE}$.

All synaptic weights satisfy the requirement that 
$$ W^{EE}>0, \quad W^{IE} > 0, \quad W^{EI}>0 $$
 where the excitatory/inhibitory character of the weight is carried by the signs in (\ref{ns1})-(\ref{ns3})

The sigmoidal transfer function $\phi(x)$ satisfies the following properties:
\begin{eqnarray*}
0<\phi(x)< 1, \quad x\in\mathbb{R}, \quad \phi'(x)>0, \quad x\in \mathbb{R}, \quad \lim_{x\rightarrow-\infty} \phi(x) = 0, \quad \lim_{x\rightarrow\infty} \phi(x) = 1 
\end{eqnarray*}
In numerical simulations, we will primarily consider the logistic transfer function
\begin{eqnarray*}
\phi(x) = \frac{1}{1+\exp(-ax)}
\end{eqnarray*} 

To prevent individual nodes from saturating due to excessively high inputs, or becoming quiescent due to insufficient input, we assume that the input weights of each neural region are normalized (Figure \ref{F0}B).  Mathematically, this assumption becomes the following:
\begin{eqnarray}
\sum_{j=1}^N \bm W^{EE}_{ij} &=& W^E, \quad i = 1,2,\ldots N, \quad \text{(Constant row some for all nodes)}\label{c1}\\  
 W^{EE}_{ij}&\geq 0&,\quad i,j = 1,2,\ldots N  \label{c2} 
\end{eqnarray}
The two conditions (\ref{c1})-(\ref{c2}) immediately imply that the weight matrix $\bm W^{EE}$ has identical $L_1$ normalized row-sums.  Recall that the $L_1$ norm of a vector $x\in \mathbb{R}^n$ is defined as $\|x\|_{1} = \sum_{i=1}^n|x_i|$.   Further, equation (\ref{c1}) also implies that the largest eigenvalue of $\bm{W^{EE}}$ is $W^E$ by the Perron-Frobenius theorem.

Before we proceed further we note that it is convenient to write the coupling matrix $\bm W^{EE}$ as follows:
\begin{eqnarray}
\bm W^{EE} = W^E \bm L^{EE}, \quad \sum_{j=1}^N \bm{L}^{EE}_{kj} = 1, \quad k = 1,2,\ldots N  
\end{eqnarray}

\subsection{Local Analysis of Equilibria } 

Before we consider the synchrony of solutions, we must first investigate when non-equilibrium solutions to (\ref{ns1})-(\ref{ns3}) actually exist.  For any connectome, the stable equilibrium point determined by the homeostatic set point destabilizes for a sufficiently large $W^E$.  This is independent of the connectome and only depends on the dynamics of the nodes.  This results in a Hopf bifurcation curve $W^{E} = g^{Hopf}(W^{IE})$ where $g^{Hopf}(x)$ can be analytically determined (see Appendix, \cite{wiltenchaos}).  This bifurcation curve has no dependence on any $L_1$ normalized connectome whatsoever.  

This Hopf bifurcation coincides with the bifurcation curve of the single recurrently coupled node.  In fact, all invariant sets (equilibria, limit cycles, etc.) for the single recurrently coupled node exist for any $L_1$ normalized connectome. For the single recurrently coupled node, these invariant sets are attractor states which starting with low amplitude oscillations, chaos, mixed-mode chaos, and mixed-mode oscillations 
(Figure \ref{F0}C-E) for progressively large $W^E$ \cite{wiltenchaos}.  As we will see from the Master Stability Function Analysis below, networks with $L_1$ normalized connectomes can synchronize to these states depending on the eigenvalue spectrum of the $L_1$ normalized weight matrix, and in particular, the magnitude of the Second Largest Eigenvalue (SLE) of the normalized connectome.  

\subsection{Master Stability Function Analysis}

To investigate this system more generally than in \cite{wiltenchaos} and determine the stability of synchronized solutions, we conducted large scale simulations of a variety of connectome types (Figure \ref{F1}).  First, as in \cite{wiltenchaos}, we found that networks of homeostatically coupled Wilson-Cowan nodes readily synchronized to attractor states of the single recurrently coupled node system with a self coupling strength of $W^E$ (Figure \ref{F1}A-D): 
\begin{eqnarray}
 \tau_1 \frac{d{E}}{dt} &=& -E + \phi\left(W^{E} E - W^{EI} I\right)\label{sn1} \\
\frac{d{I}}{dt} &=& -I + \phi\left(W^{IE} E\right) \\ 
\tau_2\frac{d{W}^{EI}}{dt} &=& I(E-p) \label{sn3} 
\end{eqnarray}
We refer to the non-equilibrium point attractor solutions to (\ref{sn1})-(\ref{sn3}) as $(E_s(t),I_s(t),W^{EI}_s(t))$.  This result was consistent across a
variety of common coupling types such as small ring networks, lattices, small world networks.   However, some connectomes could display non-synchronized solutions for a given value of $W^E$ such as larger rings ($N>9$), weak coupling networks, and larger lattices (Figure \ref{F1}E).  Thus, some connectomes have the capability to destabilize the synchronized solution arising from the single recurrently couple node. 

To investigate the loss of stability as a function of the underlying weight matrices, we slightly modified the traditional MSF approach (See Appendix) (Figure \ref{F2}A).  By linearizing around the synchronized solution ($E_s(t),I_s(t),W^{EI}_s(t))$ and assuming that $\bm W^{EE}$ is diagonalizable, one can derive the following diagonalized variational equation for perturbations to the synchronized solution:
\begin{eqnarray}
\tau_1 \frac{d\eta^{\epsilon}_k}{dt}&=&-\eta^\epsilon_k + \phi'(W^EE_s-W^I_sI_s)(\hat{r}_k \eta^\epsilon-W^{EI}_s \eta^i_k- I_s \eta^\omega_k) \label{vn1}\\
\frac{d\eta^i_k}{dt} &=& -\eta^i_k + \phi'(W^{IE} E_s) \theta \eta^i_k \\ 
\tau_2 \frac{d\eta^\omega}{dt} &=&(E_s-p)\eta^i + I_s \eta^\epsilon \label{vn3} 
\end{eqnarray}
where $\hat{r}_k$ is an eigenvalue of $\bm W^{EE}$.  

Note that for all subsequent simulations and plots, we will consider ${r}_k =\frac{\hat{r}_k}{W^E}$, which corresponds to an eigenvalue of the $\bm L^{EE}$, which has a maximal eigenvalue of 1, rather than the eigenvalues of $\bm W^{EE}$

While equation (\ref{vn1})-(\ref{vn3}), may not seem particularly useful at first glance, it can compute the stability of the synchronized state for any connectome with a simple numerical algorithm \cite{MSF} by decoupling the computation of $3N$ eigenvalue problems into $N$ separate 3 dimensional systems.  One exploits this by treating the equations (\ref{vn1})-(\ref{vn3}) as a general form and using them (in conjunction with (\ref{sn1})-(\ref{sn3})) to compute the Lyapunov exponents over a mesh in the $(\text{Re}(r),\text{Imag}(r))$ space (Figure \ref{F2}B).   Then, the eigenvalues can be computed for any connectome $\bm W^{EE}$ with the Lyapunov exponents of the synchronized solution immediately determined simply by a look-up operation on the pre-computed mesh (Figure \ref{F2}B).   

 We applied this procedure and found that for values of $W^E$ near the Hopf bifurcation of the single recurrently coupled node, the maximum Lyapunov exponent is positive for sufficiently large eigenvalue magnitudes, $|r|$.  However, for a suitably large $W^E$, the synchronized solution is stable for all $|r|$ as all Lyapunov Exponents are non-positive for $|r|<1$.    

To test this result, we considered two diagonalizable connectomes, the ring and lattice solutions, and determined when the synchronized state loses instability as a function of the network size, $N$.  Our MSF analysis shows that the eigenvalues of the ring cross into the positive Lyapunov exponent region for $N=9$ (Figure \ref{F2}C).  Indeed, we find that simulations of rings for $N=8,9,10$ shows that the synchronized solution is stable for $N=8$, and unstable for larger $N$.  However, the attractor state for $N>8$ node ring systems has periods of synchrony interspersed with bursts of asynchrony (Figure \ref{F2}D,E). For lattice-based connectomes, we find that for an $\sqrt{N}\times \sqrt{N}$ lattice, the MSF analysis predicts the instability of the synchronized solution initiating at $\sqrt{N}=16$ (Figure \ref{F2}G), which was also confirmed by our numerical simulations (Figure \ref{F2}H).  

Thus, for the Wilson-Cowan system considered here, the stability of the synchronized solution is largely determined by the SLE of the connectome.  The larger the magnitude of the SLE, the greater the potential for desynchronization, while the smaller the magnitude of the SLE, the greater the stability of the synchronous solution.

\subsection{Network Simulations with Heterogeneous Nodes} 
While MSF analysis can determine when synchronized solutions lose stability, it is limited by the assumption that all nodes are homogeneous \cite{BSF1}.  However, recent analysis of alternate diffusively coupled neural systems has indicated that so long as the heterogeneity is suitably small, MSF analysis is still a reasonable 
predictor of synchronization\cite{het1}.   

To investigate if the stability of the synchronized solution was robust to heterogeneity in the nodes, we simulated the ring networks with varying levels of heterogeneity (Figure \ref{F3}).  In particular, we considered the case where the homeostatic set point $p$ for the different nodes was chosen from a uniform distribution over small intervals around the default value of $p=0.2$:
\begin{eqnarray}
\tau_2\frac{d{W}^{EI}_k}{dt} = I_k(E_k - p_k), \quad p_k \in U[0.2-\epsilon,0.2+\epsilon] \label{het2}
\end{eqnarray}
The plastic weight $W^{EI}$ now tries to drive the excitatory nodes to their distinct set points: $E_k\rightarrow p_k$ (Figure \ref{F3}A).  

To quantify the magnitude of synchrony, we computed the Kuramoto order parameter (Materials and Methods) 
\begin{eqnarray}
R(t) = \frac{1}{N}\sum_{k=1}^N e^{i\phi_k(t)}= |R(t)|e^{i\psi(t)}
\end{eqnarray}
where $\phi_k$ is the phase of the $k$th node, as measured by $E_k(t)$ (Materials and Methods). Fully synchronized states correspond to $|R(t)|=1$ while states with a uniform (or uniformly clustered) phase distribution over the unit circle correspond to $|R(t)|=0$ (Figure \ref{F3}B).  

For small amounts of heterogeneity $(\epsilon =10^{-2}$), the stability of the synchronized solution is largely similar to the homogeneous case (Figure \ref{F3}C), with the destabilization of synchrony at a ring size of $N=9$.   However, larger amounts of heterogeneity between nodes $(\epsilon = 10^{-1})$ can completely destroy the synchronized attractor state, including restoring the stability of the equilibrium point (Figure \ref{F3}D).  Thus, the numerical simulations indicate that the synchronized solution in a mildly heterogeneous system may retain the stability characteristics of the homogeneous system, as in the network considered by \cite{het1}.

\subsection{The Second Largest Eigenvalue of Normalized Structural Connectomes}

The analysis of the model above indicates that the Second Largest Eigenvalue (SLE) of an $L_1$ normalized connectome determines the stability of the synchronized solution.  Larger values lead to instability while smaller values indicate greater stability of synchronized solutions.   This is markedly different from stability criteria invoked by the Laplacian, where the spread of the eigenvalues is often used as the determinant of stability \cite{DB1,MSF2} (Materials and Methods).  Indeed, the eigenvalues of the Laplacian and an $L_1$ normalized matrix need not coincide (Appendix), thus they generate different metrics for synchronizability that need not be related. 

Given the lack of correspondence between these synchronization criteria, we investigated whether or not the SLE differs across real connectomes (Figure \ref{F4}).   Thus, we utilized publicly available data sets from the UCLA multimodal connectivity database (NKI RS sample) consisting of DTI connectomes from 196 subjects ranging from 4-85 years of age \cite{brown}.  These connectomes consist of 188 nodes corresponding to different brain regions (Figure \ref{F4}A).  For the purposes of comparison, we considered both the $L_1$ normalization and the Laplacian transformation of the raw-connectomes (Figure \ref{F4}B). 

After normalization, we found that the SLE of the connectomes exhibited a linear decrease with age (Figure \ref{F4}C-D) as measured by a Pearson-Correlation coefficient ($\rho=-0.2812$, $p=6.54\cdot 10^{-5}$).  This decreasing strength in the SLE indicates a broad increase in the synchronizability associated with old age.  We found that this was also present in the synchronizability metric of the Laplacian, albeit with a weaker correlation ($\rho = 0.21$, $p=2.6\cdot 10^{-3}$). 

Our analysis has shown that the SLE is a potential predictor for synchronizability.  However, it is inherently global and does not implicate any particular node in the network as contributing towards synchronizability.  To investigate the impacts of individual nodes further, we applied a node-deletion protocol.  Nodes were individually deleted for each of the $n=196$, then the resulting weights renormalized and the SLE recomputed (Figure \ref{F4}F-I).   We found that across the subjects, many nodes consistently increased/decreased the SLE for the resultant matrix (Figure \ref{F4}G-H), with the most impactful nodes increasing or decreasing the SLE by no more than $1-2\%$.   We found that overall, the deletions were slightly more likely to result in SLE increases (56.46\% of deletions), rather than decreases (43.54\% of deletions), which was statistically significant (Wilcoxon Sign-Rank test, $n=36847$, $p\ll 10^-4$).  

\section{Discussion} 

With the glut of neuroimaging data, a natural question emerges as to how to best make use of structural connectomes.  We considered a dynamical systems model in the form of a network of Wilson-Cowan nodes with intranode inhibitory homeostatic plasticity, with the coupling between nodes determined by arbitrary $L_1$ normalized connectomes.  We considered a variety of connection matrix types, both synthetic (rings, weak coupling, lattices, etc.) and experimentally obtained from Diffusion Tensor Imaging data.  Through modification of the Master Stability Function (MSF) approach, we found that networks with a sufficiently large second largest eigenvalue (SLE) could destabilize the synchronized state. This result persisted when we considered small amounts of heterogeneity in the homeostatic set points of the individual nodes, but perished for a sufficiently large distribution of heterogeneity.   Finally, we tested the SLE of real-connectomes from the NKI Rockland data set and found a decrease in the SLE of the normalized-connectomes with age, indicating an increase in the network synchronizability.  

Our results indicate a general increase in synchronizability with age.  This, at face value, seems to contradict the decrease in synchronizability recently reported by \cite{DB1}.  However, there is one critical difference between the data sets we consider: the age of the subjects.  In \cite{DB1}, the authors considered subjects in the 8-22 age group and found a general decrease in synchronizability with age in 882 developing subjects.  Our result, on the other hand, consists of 196 subjects over a much broader age range (7-85), and we find an increase, rather than a decrease in synchronizability with old age.  One possible parsimonious explanation is that both results are valid.  In developing children, there is a general decrease in synchronizability which reaches a plateau.  However, as adults enter into old age, synchronizability begins to increase again due to white matter degeneration.  Thus, these combined results would indicate that for a sufficiently large sample covering a broad age range, synchronizability is a unimodal function with a minimum at adulthood.  This hypothesis is supported by the epidemiology of epilepsy.  The prevalence of epilepsy with age is a u-shaped curve with increased synchronizability in children and the elderly  \cite{epilepsy1,epilepsy2}.  

Recently, the applicability of MSF analysis to brain synchronizability has been criticized \cite{BSF1}.  Two of the major criticisms levied at MSF analysis in \cite{BSF1} are 1) The functional form of the MSF is dependent on the dynamics of the system, thus considering the eigenvalues alone can lead to erroneous results, 2) the homogeneity and diffusive coupling assumption in the MSF approach make it difficult to apply.  All of the criticisms raised by \cite{BSF1} are indeed valid.  Here, we find that by considering a Wilson-Cowan system with homeostatically regulated inhibitory plasticity yielded one of two behaviours:  Stability of the synchronized state regardless of the eigenvalues of the normalized matrix, or instability of the synchronized solution for weight matrices with a sufficiently large SLE.   Unlike prior work (\cite{DB1}), we considered the normalized coupling case, yielding alternate MSF profiles and biologically plausible modelling for the neural regions.  Further, we did find that the MSF analysis was somewhat permissive of heterogeneity in the nodes.   While the criticisms in \cite{BSF1} are valid, the analyses conducted here and in \cite{DB1} do yield diagnostic measures that may be worth examining in real subjects for comparison across subjects.  


\clearpage


\section{Figures}
\begin{figure}[htp!]
\centering
\includegraphics[scale=0.40]{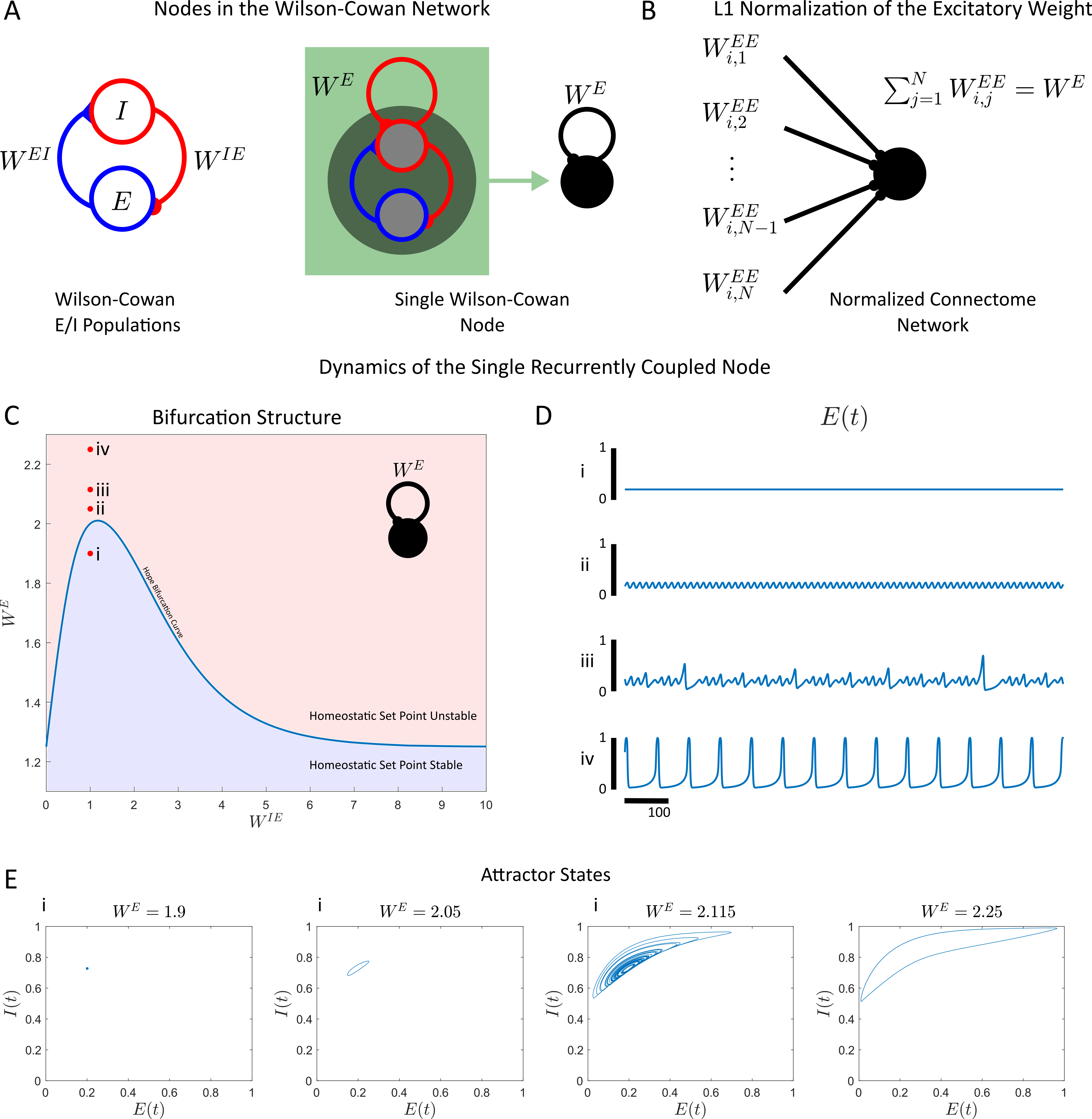}
\caption{$L_1$ Normalization and Dynamics of the Single Recurrently Coupled Node} \label{F0}
\end{figure}

\clearpage
\subsection*{Figure 1:  Large Scale Synchronization to The Single Self-Coupled Node Dynamics} 
\textbf{(A)}  A single node of the Wilson-Cowan system (left) with homeostatic plasticity consists of an excitatory population (E) an inhibitory population (I) and weights $W^{IE}$, $W^{EI}$ coupling these two populations together.  The weight $W^{EI}$ is dynamic and serves to maintain a homeostatic set point in $E$.  Nodes are connected together via the excitatory populations with weights $W^{EE}$ (middle). An important driver of network behaviour is the single recurrently coupled node (right).   \textbf{(B)}. The normalization condition applied in larger networks, $\sum_{j=1}^NW^{EE}_{ij} =W^E$ is an $L_1$ normalization condition on each vector of input weights.  \textbf{(C)} The local bifurcation diagram of the single recurrently coupled node, a Hopf bifurcation curve (blue) delineates the region where the homeostatic set point is stable (blue), and unstable (red). \textbf{(D)}   Dynamics of the excitatory population for different values of $W^E$: $W^E = 1.9$ (i), $W^E = 2.05$ (ii), $W^E = 2.115$ (iii), $W^E=2.25$ (iv).  \textbf{(E)} Projections of the steady state attractors into the (E,I) plane.  

\clearpage

\begin{figure}
\centering
\includegraphics[scale=0.65]{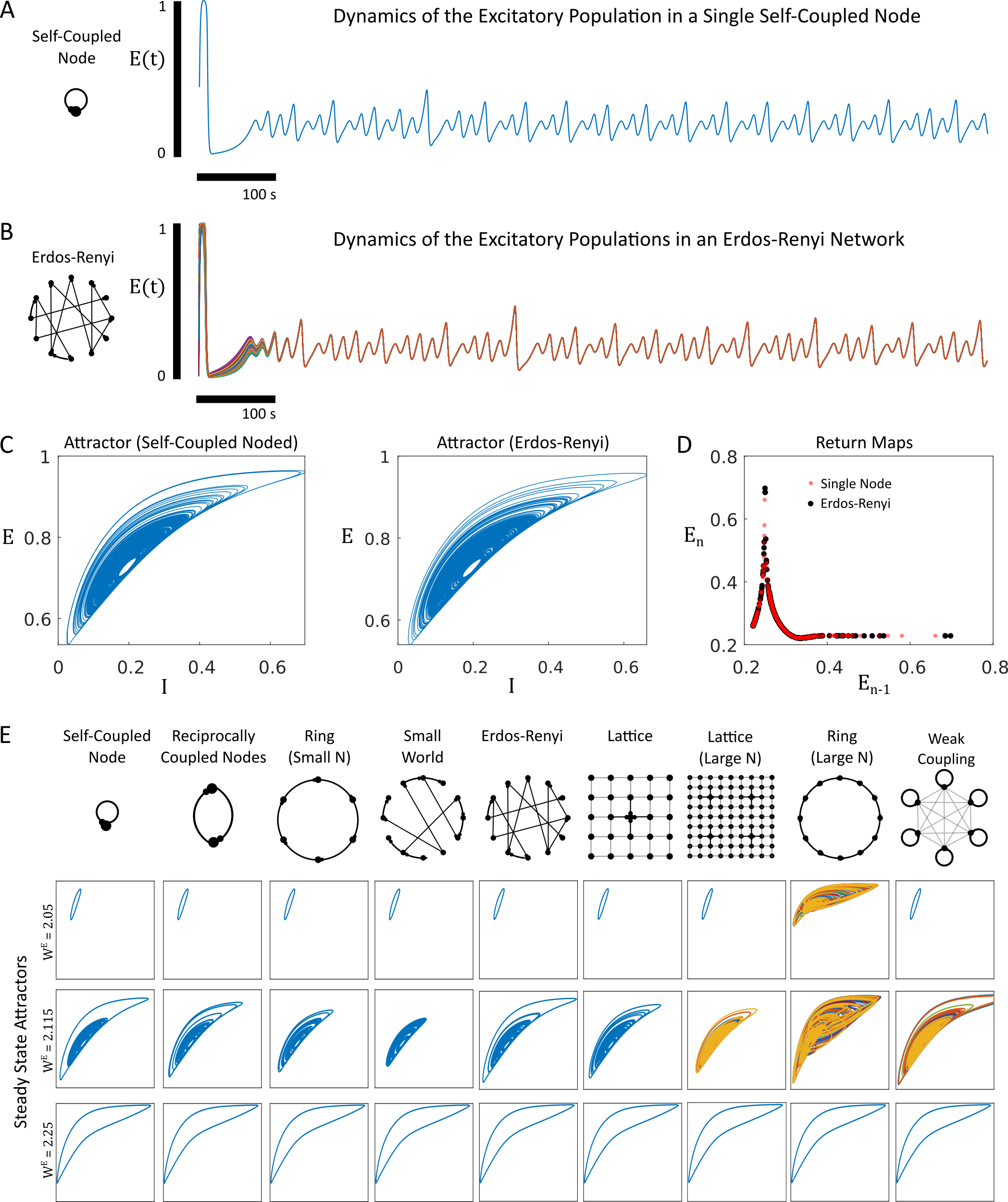}
\caption{Large Scale Synchronization to The Single Self-Coupled Node Dynamics} \label{F1}
\end{figure}

\clearpage
\subsection*{Figure 2:  Large Scale Synchronization to The Single Self-Coupled Node Dynamics} 
\textbf{(A)}  The activity for the excitatory population $E(t)$ for a single recurrently coupled node for $W^E = 2.115$, in the chaotic dynamics regime considered by \cite{wiltenchaos}.   \textbf{(B)} An Erd\H{o}s-R\'enyi coupled network with $N=100$ nodes, with an $L_1$ normalized connectome with an identical $W^E$ as (A).  The nodes all synchronize onto a chaotic attractor.  \textbf{(C)} The steady state attractors for the single-node system in (A) and the 100 node Erd\H{o}s-R\'enyi system in (B).  \textbf{(D)} The return maps for the excitatory populations of both networks, where the $(n-1)$th peak of $E(t)$ is plotted against the $n$th peak.  As the return maps broadly overlap, the two attractors are identical.  \textbf{(E)} Steady state attractors for 1) a single self coupled node at $W^E = 2.05$, $W^E = 2.115$, and $W^E = 2.25$, two nodes with reciprocal coupling, a ring consisting of $N=7$ nodes, A Watts-Strogatz small-world network with $N=200$ nodes, an Erd\H{o}s-R\'enyi Network with $N=100$ nodes, a $10\times 10$ lattice network, a $20\times20$ lattice network, a ring with $N =10$ nodes, a system with $N=50$ nodes that have strong self-coupling, but random weak-coupling to other nodes.  Some systems display a robust synchronization phenomenon across states, while other systems desynchronize.  

\clearpage

\begin{figure}
\centering
\includegraphics[scale=0.6]{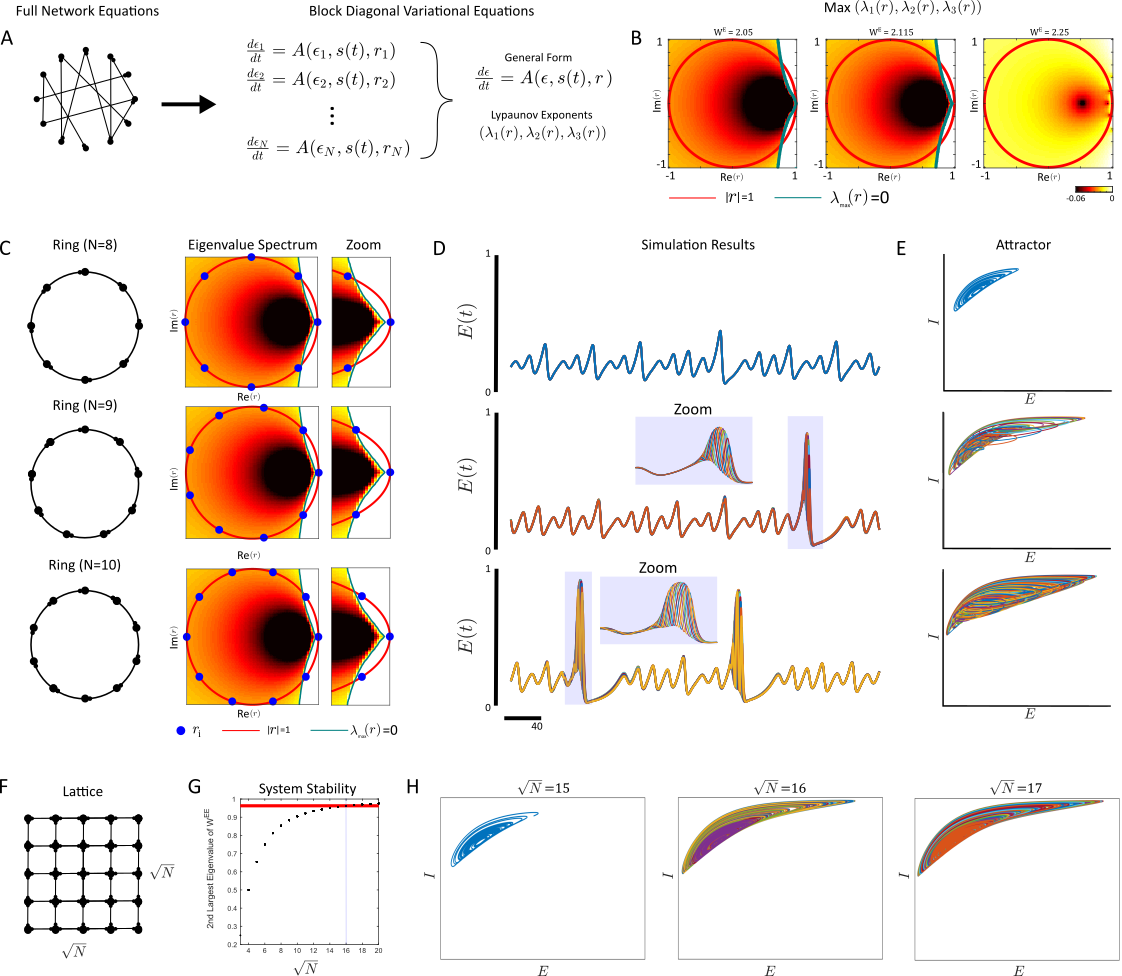}
\caption{Master Stability Function Analysis for $L_1$ Normalized Connectomes} \label{F2}
\end{figure}

\clearpage
\subsection*{Figure 3:  Master Stability Function Analysis for $L_1$ Normalized Connectomes} 
\textbf{(A)}  To derive the master stability function for the synchronized solution (SS) to any connectome, the original dynamical system is first linearized and diagonalized into a general form involving the independent perturbations to the dynamics $\epsilon$, and the eigenvalues of the matrix $\bm W^{EE}$.   The master stability function can be numerically approximated for any weight matrix by resolving the Lyapunov exponents for the general block over a mesh in $r$.  The largest positive eigenvalue yields the master stability function.  \textbf{(B)}  The numerically derived master-stability functions for the WC system (\ref{ns1})-(\ref{ns3}) for the values $W^E = 2.05$ (limit cycle SS), $2.115$ (chaotic attractor SS), $2.25$ (relaxation oscillator SS).   \textbf{(C)} The master stability function for $W^E = 2.115$ applied to rings of increasing size.  \textbf{(D)} The dynamics of simulated network.  Note that the non-fully synchronized attractors for $N=9,10$ display periods of synchronization along with periods of desynchronization.   \textbf{(E)} Steady state attractors projected onto the $(E,I)$ plan for the $N$ node ring network.     \textbf{(F)}  Lattices of size $\sqrt{N}\times \sqrt{N}$ with nearest neighbour coupling.   \textbf{(G)} The master stability analysis for $W^E=2.115$ predicts the onset of instability to the synchronous solution at $\sqrt{N}=16$.   \textbf{(H)}  Network attractor states for $\sqrt{N}=15,16, 17$.  
\clearpage

\begin{figure}
\centering
\includegraphics[scale=0.7]{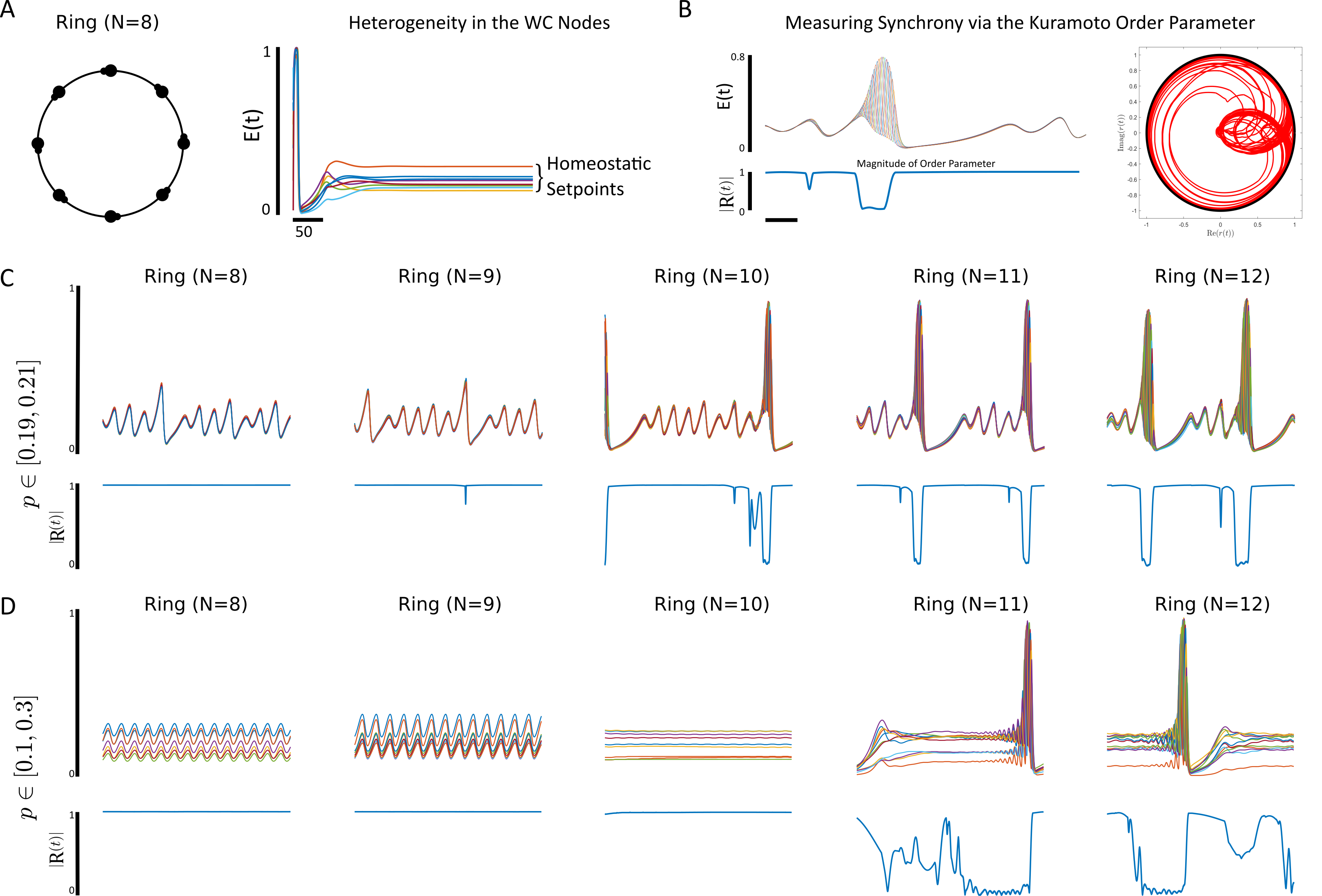}
\caption{Synchronization Persists for Small Amounts of System Heterogeneity} \label{F3}
\end{figure}

\clearpage
\subsection*{Figure 4: Synchronization Persists for Small Amounts of System Heterogeneity} 
\textbf{(A)}  
An identical ring network as in Figures \ref{F1} and \ref{F2} only with different homeostatic set points for each single node, and $W^E = 2.115$.  \textbf{(B)}  To quantify partial synchrony in heterogeneous systems, we employed the Kuramoto order parameter $R(t)$.  Perfectly synchronized systems have an $|R(t)|=1$ while asynchronously distributed systems have $|R(t)| = 0$   
\textbf{(C)} Simulated networks (top) of $N=8-12$ rings with heterogeneity in their homeostatic set points ($p$ randomly distributed with a uniform distribution in $[0.19,0.21]$).  The magnitude of the Kuramoto order parameter is plotted in blue (bottom).   
\textbf{(D)} Identical to (C), only with $p$ uniformly distributed in $[0,1,0.3]$ 

\clearpage

\begin{figure}
\centering
\includegraphics[scale=0.24]{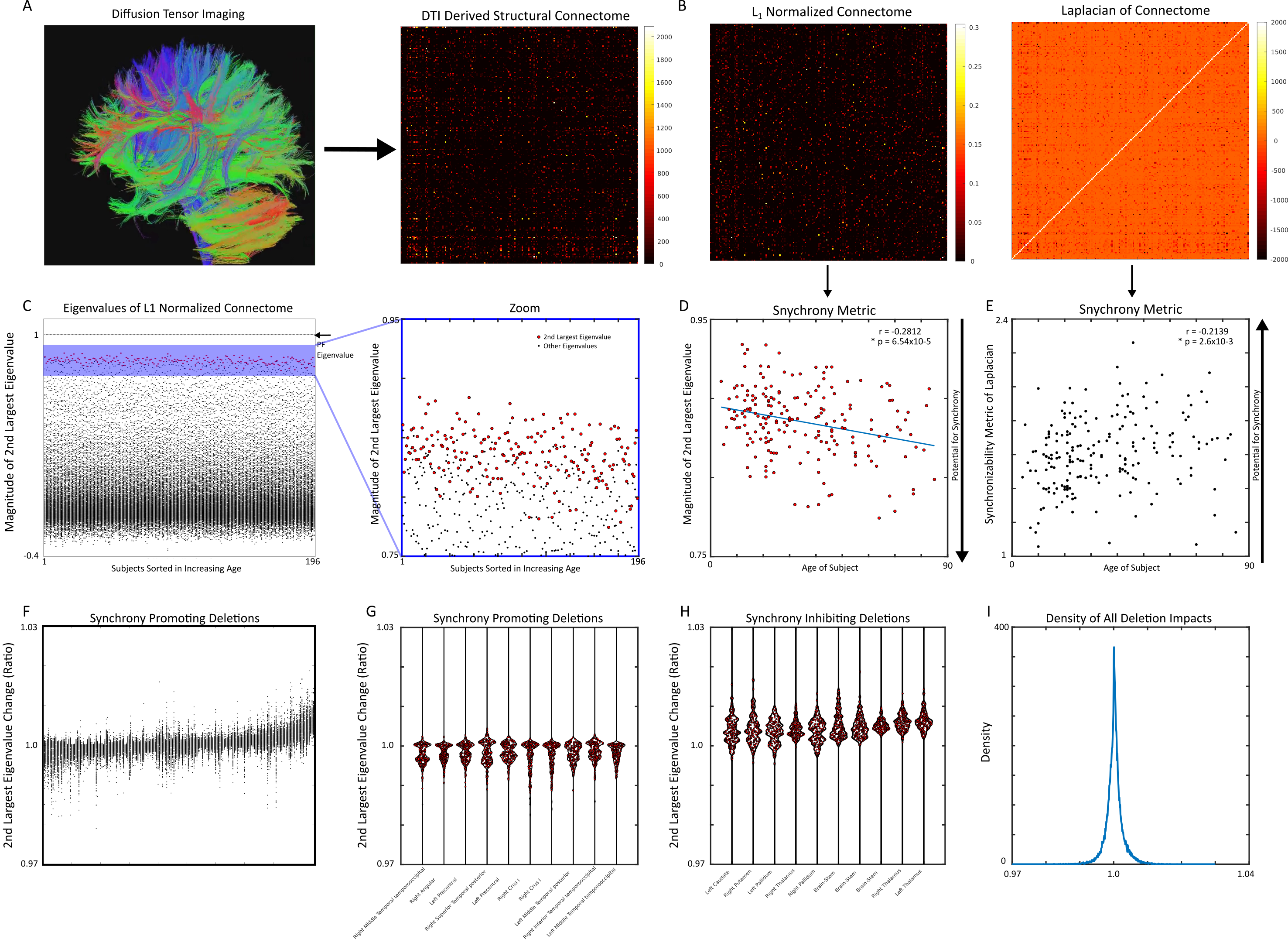}
\caption{The Second Largest Eigenvalue of $L_1$ Normalized Connectomes Decreases with Age } \label{F4}
\end{figure}

\clearpage
\subsection*{Figure 5: 
The Second Largest Eigenvalue of $L_1$ Normalized Connectomes Decreases with Age } 

\textbf{(A)} Diffusion Tensor Imaging (DTI) is used to create a connectivity matrix detailing the white-matter tracts connecting different regions of the brain.  \textbf{(B)} The raw DTI connectivity matrix can be transformed into matrices which allow us to apply master stability function analysis:  either the $L_1$ normalized connectome (normalized coupling) or the Laplacian (diffusive coupling) 
\textbf{(C)}  The eigenvalues of $L_1$ normalized connectivity matrix.  The dominant eigenvalue is always the Perron-Frobenius (PF) eigenvalue ($r_{max}=1$).  The second largest eigenvalue is highlighted (red) (Zoom, right).  \textbf{(D)}  The second largest eigenvalue decays with age $r=-0.2812$, $p = 6.54\times 10^{-5}$.  \textbf{(E)} The synchronizability metric for the Laplacian (see Materials and Methods).  \textbf{(F)}  Nodes are sequentially deleted while the second largest eigenvalue of the renormalized connectome is recomputed and compared to the full weight matrix.     
 \textbf{(G)} The top 10 synchrony promoting deletions, as measured by smallest relative change of the second largest eigenvalue.  \textbf{(H)}  The top 10 synchrony inhibiting deletions, as measured by the largest relative change of the second eigenvalue.  
\textbf{(I)}  The distribution of all deletions. 

\clearpage

\section*{Acknowledgements} 

WN is funded by an NSERC Discovery Grant, and a Hotchkiss Brain Institute start-up fund.  SAC is funded by an NSERC Discovery Grant. We would like to thank Joern Davidsen for many fruitful conversations about this work.

\section{Methods}

All simulations were conducted in MATLAB2019a, with the ODE45 numerical integration sub-function used for all simulations.  The tolerance parameter values (RelTol and AbsTol) were increased to $10^{-14}$ to ensure accurate integration from the default $10^{-6}$ values.

\subsection{Network Sub-type Parameters}

For all networks considered, the parameters used were as in Table \ref{T1}. 
\begin{table}[htp!]
\centering
\begin{tabular}{ |l |c|}
\hline
  $\tau_1$ & 2  \\
\hline
   $\tau_2$ & 5  \\
\hline
  $a$ &  5 \\
\hline
 $p$ & 0.2 \\
\hline
$N$ (Ring) & 7 (Small), 10 (Large)\\ 
\hline
$N$ (Erd\H{o}s R\'enyi)  & 100 \\ 
\hline
$N$ (Small World)  & 200\\ 
\hline
$N$ (Lattice) & 100 (Small), 400 (Large) \\
\hline
$N$ (Weak Coupling) & 50  \\
\hline
\end{tabular}
\caption{Table of parameters for simulations, unless otherwise specified by a figure caption.} \label{T1}
\end{table}

\subsubsection{Ring Networks} 

Rings were generated by having each node connected to the next with a weight of $W^E$:  
$$ W^{EE}_{i+1,i} = W^{E}, \quad i=1,2,\ldots N-1, \quad W^{EE}_{1,N} = W^E$$

\subsubsection{Small World Networks} 
Small world networks are generated by creating a bidirectional ring network initially where neuron is connected to its $k$ previous and next nearest neighbours with the weight $\frac{W^E}{2k}$.  Then, a parameter $0\leq\beta\leq 1$ is used to set the probability of random rewirings where node $(i,j)$ is randomly permuted with $(i,k)$ for each $i=1,2,\ldots N$, and $i<j<i+k/2$.   The values $\beta = 0.7$ and $k=20$ were taken in all simulations.

\subsubsection{Erd\H{o}s R\'enyi Networks} 
Erd\H{o}s R\'enyi networks were generated by first creating a random weight matrix $Q^{EE}_{ij} = q_{ij}$ where $q_{ij}$ was a uniform random variable on $[0,1]$, then normalizing rows in $Q^{EE}$ with $W^{EE}$.

\subsubsection{Lattice Networks} 

Lattices are generated by setting the nodes on a $\sqrt{N}\times\sqrt{N}$ square lattice. Node ${i,j}$ receives input weights of strength $\frac{W^E}{4}$ from node $(i+1,j)$, $(i-1,j)$, $(i,j+1)$ and $(i,j-1)$.  
Finally, periodic boundary conditions are applied at the edges $i=N,1$ and $j=N,1$.

\subsubsection{Weak Coupling Networks} 

Weak coupling networks are generated by first creating a random matrix:
\begin{eqnarray}
{Q}^{EE}_{ij} = \delta_{ij} + \lambda\cdot q_{ij} 
\end{eqnarray}
where $q_{ij}$ is a uniformly distributed random number, and $\delta_{ij}$ is the Kronecker delta.  
Then, $\bm W^{EE}$ is given by normalizing $\bm Q^{EE}$ with $W^{E}$:
$$ W^{EE}(i,j) = W^{E}\cdot\frac{Q^{EE}(i,j)}{\sum_{j=1}^N Q^{EE}(i,j)}$$ 
The parameter $\lambda$ controls how weak the non-self-coupling components of the weight matrix were and was set to  $\lambda = 10^{-3}$.

\subsection{Numerically Computing Lyapunov Exponents} 

The Lyapunov exponents were numerically computed by the algorithm described in \cite{wolf} (see Appendix).  

\subsection{Numerically Computing the Kuramoto Order Parameter} 

The Kuramoto order parameter is computed directly by first using the findpeaks function in MATLAB2019a to detect the peaks of the chaotic or periodic solutions.  These peaks are taken to have a phase of $2\pi$ at time $t^*$, the location of the peak $(\phi_k(t^*) = 2\pi)$ with the phase value reset to 0 in the next time step $t^* + \Delta t$.  The phase is linearly interpolated for all other time points between $[0,2\pi]$.  While this is a crude approximation to the actual phase, we found that other numerical methods (e.g. the Hilbert Transform) were yielding poor estimates to the phase due to the mixed mode nature of the periodic or chaotic solutions.

\subsection{Synchronization Metric of the Laplacian} 

Following \cite{DB1,MSF2}, we also computed the synchronization metric ($\sigma^{-2}$) of the Laplacian as: 
\begin{eqnarray*}
\frac{1}{\sigma^2} &=& \frac{d^2(N-1)}{\sum_{i=1}^{N-1} (\lambda_i - \bar{\lambda})^2} \\ 
\bar{\lambda} &=& \sum_{i=1}^{N-1} \lambda_i \\\
d &=& \frac{1}{N} \sum_{i} \sum_{j\neq i} L_{ij}
\end{eqnarray*}  
where $L$ is the Laplacian of DTI structural connectivity matrix and $\lambda_i$ are the eigenvalues of the $L$.  Note that as the row-sum of the Laplacian is identically 0, only the non-zero eigenvalues (hence the $N-1$ terms in the sums) are used to compute this metric
%

\clearpage
\section*{Appendix A: Master Stability Function Derivation} 
\subsection{Appendix A1: Laplacian/Diffusive Connectivity} 

The Master Stability Function (MSF) approach introduced originally by \cite{MSF} allows one to analyze the stability of synchronized solution $\bm x_i(t) =\bm x_s(t), i=1,2,\ldots N$, of the coupled dynamical network:
\begin{eqnarray}
\dot{\bm x}_i = F(\bm x_i) + \sigma\sum_{j=1}^N \bm A_{ij}G(\bm x_j), \quad \bm x_i = (x_{i1},x_{i2}\ldots x_{ip})
\label{genmod}
\end{eqnarray}
by knowing little more than the eigenvalues of of the matrix $\bm A$. To apply the MSF formulation, one requires that $\sum_{j=1}^N A_{ij} = 0$, $\forall i$, and that 
\begin{eqnarray}
\dot{\bm x_s} = F(\bm x_s)  
\end{eqnarray}
These two criteria force the synchronized solution to be an invariant set of the coupled network.  The stability analysis is accomplished by first linearizing the subsequent dynamics around the invariant set $\bm x_s(t)$ :
\begin{eqnarray}
\dot{\bm \epsilon}_i=
DF(\bm x_s(t)) \bm \epsilon_i+ \sigma \sum_{j=1}^N \bm A_{ij} DG(\bm x_s(t))
\bm \epsilon_j, \quad \bm \epsilon_i = (\epsilon_{i1},\epsilon_{i2},\ldots \epsilon_{ip}) \quad, i=1,2,\ldots N  \label{lp1} 
\end{eqnarray}
where $DF$ and $DG$ denote the Jacobians of $F$ and $G$. This non-autonomous dynamical system can resolve the Lyapunov exponents of the synchronized solution $x_s(t)$ for any $\bm A$.  However, for diagonalizable matrices $\bm A$, i.e., 
matrices such that $\bm P^{-1} \bm A \bm P$ is diagonal, one can consider the substitution:
\begin{eqnarray}
\begin{pmatrix}\eta_{1,j}\\ \eta_{2,j}\\ \vdots\\ \eta_{N,j} \end{pmatrix} 
= \bm P^{-1} \begin{pmatrix}\epsilon_{1,j}\\ \epsilon_{2,j}\\ \vdots\\ \epsilon_{N,j} \end{pmatrix}, \quad  j =1,2,\ldots p
\end{eqnarray} 
which yields the following: 
\begin{eqnarray}
\bm \eta_i ' = \left(DF(\bm x_s(t)) + \sigma r_i DG(\bm x_s(t))\right)\bm \eta_i, \quad i =1,2,\ldots N \label{lp2} 
\end{eqnarray}
which is a decoupled block diagonal system.   This reduces the problem of determining $Np$ Lyapunov exponents in (\ref{lp1}) to solving separate $N$ separate Lyapunov exponents in $p$ dimensional dynamical systems in (\ref{lp2}).  The final step in MSF analysis is to note that (\ref{lp2})
 is a generic system of the eigenvalue $r_i$ which can be considered as a free parameter, $r$, thus yielding the master stability function $\lambda_{max}(r)$ as the maximum Lyapunov exponent of (\ref{lp2}) for a generic $r$.  

The requirement that $\sum_{j=1}^N A_{ij}=0$ is seldom satisfied by real-world connectomes, and thus users of MSFs often assume so-called ``diffusive coupling":
\begin{eqnarray}
\dot{\bm x}_i &=& F(\bm x_i) -\sigma \sum_{j=1}^N \bm A_{ij}(G(\bm x_j)-G(\bm x_i)), \quad \bm x_i = (x_{i1},x_{i2}\ldots x_{ip})\\
 &=& F(\bm x_i) -\sigma \sum_{j=1}^N \bm A_{ij}G(\bm x_j) + \sigma \left( \sum_{j=1}^N \bm A_{ij}\right) G(\bm x_i), \quad \bm x_i = (x_{i1},x_{i2}\ldots x_{ip})\\
&=& F(\bm x_i) - \sigma\sum_{j=1}^N L_{ij} G(x_j) 
\end{eqnarray}
where $L_{ij} = \delta_{ij} \cdot\left( \sum_{j=1}^N A_{ij}\right)-A_{ij}$.  This assumption of the structural form of the underlying network model allows one to use an arbitrary connectome $A_{ij}$ with MSF analysis, albeit with some rigid assumptions on the nature of the coupling.  Thus, in the diffusive coupling case, one considers an arbitrary connectome and analyzes the eigenvalues of the Laplacian, $L$, to determine the stability of the synchronized state.  

\subsection{Appendix A2: Local Stability Analysis of Equilibria}
Somewhat surprisingly, the local analysis of equilibria is largely similar for arbitrary $N$, and $L_1$ normalized $\bm L^{EE}$ \cite{wiltenchaos}.  In particular, we have the following: 

\begin{itemize}
\item[1.] All nodes share the common equilibrium point $E_k = p, I_k = \phi(\theta p), W^{EI}_k = \frac{W^E p -\phi^{-1}(p)}{\phi(W^{IE}p)}$, $k=1,2,\ldots N$.  This is the only equilibrium point of the $N$ node system.  For convenience, we will refer to this equilibrium point as $\bm z$
\item[2.]  The characteristic polynomial for the system (\ref{ns1})-(\ref{ns3}) always decomposes into a product of $N$ cubic polynomials, each cubic polynomial is of the same general form, and whose coefficients depend on the eigenvalues $r_1,r_2,\ldots r_N$ of $\bm W^{EE}$ as the following:
\begin{eqnarray}
C(\lambda) &=& \prod_{i=1}^N\left(Q(\lambda) - W^E\frac{r_i\lambda(\lambda+1)\phi'(\phi^{-1}(p))}{\tau_1} \right)\\
Q(\lambda) &=& \lambda^3 + \lambda^2\left(\frac{1}{\tau_1} + 1\right) +\lambda\left(\frac{1}{\tau_1} + \frac{\bar{W}^{EI}\phi'(\phi^{-1}(p))\phi'(\theta p)\theta}{\tau_1} + \frac{\bar{I}\phi'(\phi^{-1}(p))}{\tau_1\tau_2}\right) +\frac{\bar{I}\phi'(\phi^{-1}(p))}{\tau_1\tau_2}
\end{eqnarray}
\item[3.] The cubic polynomial associated with the maximum eigenvalue of $\bm L^{EE}$ induces a Hopf bifurcation of the equilibrium point.  While the general form for the bifurcation curve is complicated, it is typically of the form 
\begin{eqnarray}
W^{EE} = g(W^{IE}) 
\end{eqnarray} 
where $g(x)$ is a nonmonotonic, unimodal function (Figure \ref{F0}C) \cite{wiltenchaos}.  

\end{itemize}
Result 1 follows directly from the form of (\ref{ns1})-(\ref{ns3}), while Result 2-3 are derived in \cite{wiltenchaos}.  
 
\subsection{Appendix A3: Normalized Connectivity}

Here, we will apply the MSF approach to the Wilson Cowan system in (\ref{ns1})-(\ref{ns3}).  Rather than deriving an alternate form of the MSF for a general model such as \eqref{genmod}, we will proceed directly with the derivation for system (\ref{ns1})-(\ref{ns3}).  The extension to other models is straightforward. First we note that with the $L_1$ normalization condition \eqref{c1},
the synchronized solution$(E_s(t),I_s(t),W^I_s(t))$  is a solution to the model for a single recurrently coupled node:
\begin{eqnarray}
\tau_1\frac{dE_s}{dt} &=& -E_s + \phi(W^{E}E_s - W^{I}_sI_s) \label{sn1b}\\
\frac{dI_s}{dt} &=& - I_s + \phi(\theta E_s)\\
\tau_2\frac{dW^{I}_s}{dt} &=& (E_s-p)I_S \label{sn3b} 
\end{eqnarray}
and thus is an invariant set of the full $N$ node system for all connectomes with normalization constant $W^E$.  Then, consider the variational equations generated by
\begin{eqnarray}
\epsilon_k = E_k - E_s, \quad i_k = I_k - I_s,\quad \omega_k = W^I_k - W^I_s  
\end{eqnarray}
Then, we have the following:
\begin{eqnarray*}
\epsilon_k' &=& E_k' - E_s'  =
 \frac{1}{\tau_1}\left(-\epsilon_k + \phi'(W^EE_s - W^I_sI_s)\left(\sum_{j=1}^N W^{EE}_{kj}\epsilon_j -W^I_s i_k - \omega_kI_s \right) +H.O.T.\right)\\
i_k' &=& I_k' - I_s' = -i_k + \phi'(\theta E_s)\theta \epsilon_k + H.O.T. \\
\omega_k' &=& (W^I_k)' - (W^I_s)' = \frac{1}{\tau_2}\left(i_k(E_s-p) + I_s\epsilon_k)\right)+H.O.T
\end{eqnarray*}
where $H.O.T.$ denotes Higher Order Terms.  This system written in matrix form becomes: 
\begin{eqnarray}
\tau_1\bm \epsilon' &=& -\bm \epsilon+\phi'(W^E E_s - W^I_sI_s)\cdot \left(\bm W^{EE} \epsilon - W^I_s\cdot \bm i - I_s\cdot  \bm \omega \right)\\
\bm i'&=& -\bm i + \phi'(\theta E_s)\theta \bm \epsilon \\ 
\tau_2 \bm \omega' &=&  (E_s-p)\cdot \bm i+ I_s\cdot \bm \epsilon 
\end{eqnarray}
As in the traditional MSF approach, we consider the case where the matrix $\bm W^{EE}$ is diagonalizable: 
\begin{eqnarray}
\bm W^{EE} = \bm P\bm D\bm  P^{-1}, \quad \bm \eta_\epsilon =\bm P^{-1} \bm\epsilon, \bm \eta_i =\bm P^{-1}\bm i, \quad \bm \eta_\omega = \bm P^{-1}\bm \omega  
\end{eqnarray}
Then we have:
\begin{eqnarray}
\tau_1 \bm \eta_\epsilon'  &=& \bm P^{-1}\left(-\bm P\bm \eta_\epsilon+\phi'(W^E E_s - W^I_sI_s)\left(\bm P\bm D\bm \eta_\epsilon - W^I_s\cdot \bm P \bm \eta_i - I_s\cdot  \bm P\bm \eta_\omega\right)\right)\\
\tau_1 \bm \eta_\epsilon' &=& -\bm \eta_\epsilon + \phi'(W^E E_s - W^I_sI_s)\cdot \left(\bm D \bm \eta_\epsilon - W^I_s\cdot \bm \eta_i - I_s\cdot  \bm \eta_\omega \right)\\
\bm \eta_i' &=& -\bm \eta_i + \phi'(\theta E_s)\theta\cdot \bm \eta_\epsilon\\
\tau_2\bm \eta_\omega' &=& (E_s-p)\cdot \bm \eta_i +I_s\cdot \bm \eta_\epsilon
\end{eqnarray}
Which yield the following $N$ independent, 3-dimensional systems:
\begin{eqnarray}
\tau_1 \eta_{\epsilon}'&=&-\eta_\epsilon + \phi'(W^EE_s-W^I_sI_s)(\hat{r}_k \eta_\epsilon-W^I_s \eta_i - I_s \eta_\omega)\\
\eta_i' &=& -\eta_i + \phi'(\theta E_s) \theta \eta_\epsilon \\ 
\tau_2 \eta_\omega' &=&(E_s-p)\eta_i + I_s \eta_\epsilon 
\end{eqnarray}

The important point here is that for some attractor state for fixed $W^E$, $(E_s(t),I_s(t),W^I_s(t))$, we can assess the stability for any weight matrix $\bm W^{EE}$ by analyzing the stability of the equilibrium point of the decoupled blocks $\bm z$ above.  These blocks only differ based on the dependence of the eigenvalues $\hat{r}_k$ of $\bm W^{EE}$.  In fact, for fixed $W^E$ we need only do the analysis for $\hat{r}_k$ such that $|r_k|\leq W^E$, as required by the $L_1$ normalization constraint on $W^{EE}$.   This system in conjunction with the system (\ref{sn1b})-(\ref{sn3b}) is used to compute the Lyapunov spectrum for any possible normalized connectome by varying $\hat{r}_k$ over a 2D mesh, and computing the maximum Lyapunov exponent over this mesh.

\subsection{Appendix A4: Lack of Correspondence Between Eigenvalues of Different Matrix Transforms} 

While the raw-DTI matrix eigenvalues are fixed, one potentially reasonable assumption is that the eigenvalues of the $L_1$ normalized connectome and the Laplacian are somehow related. Here, we will show with counterexamples that this is not the case.  In particular, consider the $2\times2$ symmetric matrix

\begin{eqnarray*}
D = \begin{pmatrix} w_1 & w_2 \\ w_2 & w_3 \end{pmatrix}, \quad L(D) = \begin{pmatrix} w_2 & -w_2 \\ -w_2 & w_2 \end{pmatrix}, \quad N(D) = \begin{pmatrix} \frac{w_1}{w_1+w_2} & \frac{w_2}{w_1+w_2} \\ \frac{w_2}{w_2+w_3} & \frac{w_3}{w_2+w_3} \end{pmatrix}
\end{eqnarray*}
which yields as eigenvalues 
\begin{eqnarray*}
\lambda(L(D)) = \{0, 2w_2\}, \quad \lambda (N(D)) = \left\{1, \frac{w_3w_1 - w_2^2}{(w_1+w_2)(w_3+w_2)}\right\} 
\end{eqnarray*}
Thus, the eigenvalues of the Laplacian are entirely determined (for this case) by the off diagonal elements, while the eigenvalues of the normalized matrix are a rational function dependent on all matrix elements.

\end{document}